\documentclass[11pt]{article}

\usepackage{graphicx}
\usepackage{latexsym}
\usepackage{amsmath,amsthm,amssymb,amscd}
\usepackage{epsf,amsmath}
\newtheorem{lemma}{Lemma}

\newtheorem{theorem}{Theorem}
\newtheorem{proposition}{Proposition}

\numberwithin{assumption}{section} \numberwithin{lemma}{section}
\numberwithin{definition}{section} \numberwithin{theorem}{section}
\numberwithin{proposition}{section} \numberwithin{remark}{section}
\numberwithin{corollary}{section} \numberwithin{example}{section}
\numberwithin{conclusion}{section} \numberwithin{equation}{section}
\numberwithin{remark}{section} \numberwithin{table}{section}

\def\var{{\rm Var}}

\def\N{{\rm N}}

\def \I{\rm I}

\newcommand{\va}{\mbox{\boldmath $a$}}

\newcommand{\vb}{\mbox{\boldmath $b$}}

\newcommand{\vw}{\mbox{\boldmath $w$}}

\newcommand{\vz}{\mbox{\boldmath $z$}}

\newcommand{\vy}{\mbox{\boldmath $y$}}
\newcommand{\vx}{\mbox{\boldmath $x$}}

\newcommand{\vbeta}{\mbox{\boldmath $\beta$}}

\newcommand{\vxi}{\mbox{\boldmath $\xi$}}

\newcommand{\vmu}{\mbox{\boldmath $\mu$}}
\newcommand{\vepsilon}{\mbox{\boldmath $\epsilon$}}

\newcommand{\vone}{\mbox{\boldmath $1$}}

\begin{document}

\begin{center}
{\Large\bf  
Sequential Lasso for feature selection with ultra-high dimensional feature space \\}

\vspace{0.15in}

{\sc By}  SHAN LUO${}^{1}$ {\sc and}  ZEHUA CHEN${}^{2}$\\
\vspace{0.15in}

${}^{1,2}$Department of Statistics and Applied Probability \\
National University of Singapore \\
${}$ \\
Email: ${}^{1}$luoshan08@nus.edu.sg,   ${}^{2}$stachenz@nus.edu.sg. 
\vspace{1in}

% Running title:   Sequntial Lasso
\end{center}

\begin{abstract}
We propose  a novel  approach, Sequential Lasso, for feature selection 
 in linear regression models with ultra-high dimensional feature spaces.   We investigate  in this article the asymptotic properties of Sequential Lasso and establish its selection consistency. Like other sequential methods, the  implementation of Sequential Lasso is not limited by the dimensionality of the feature space. It has advantages over other sequential methods. The simulation studies  comparing Sequential Lasso with other sequential methods are reported. .  

 \end{abstract}

\newpage
\section{Introduction}

The so-called small-$n$-large-$p$ problems are abundant in many important contemporary scientific fields.  A small-$n$-large-$p$ problem refers to the situation where the number of covariates  is huge,  though only a few of them  are causally related to the response variable under study, but the sample size is relatively small.  There are two related but different purposes in the study of small-$n$-large-$p$ problems: (i) to build a model with good prediction properties and (ii) to identify the covariates  which are causally related to the response variable.  The classical least square regression approach is no longer appropriate in small-$n$-large-$p$ problems because of the sheer huge number of the covariates.  Feature selection becomes crucial  (by a feature we mean a covariate or a function of covariates such as the product of any two covariates, etc.).

Since the seminal paper on LASSO \cite{Tibshirani 1996} published in 1996,  a great interest has been focused on penalized likelihood methods.  These methods include SCAD \cite{Fan and Li 2001}, Elastic net \cite{Zou and Hastie 2005},  Adaptive LASSO \cite{Zou2006},  Bridge \cite{FrankFriedman1993}, etc., to name but a few. A penalized likelihood method selects variables and estimates the coefficients at the same time.  Under certain conditions, the various penalized likelihood methods have the so-called oracle property; that is,  asymptotically, the set of causal variables can be identified exactly and the estimated coefficients are $L_2$ consistent, see
\cite{Fan and Li 2001}  \cite{Jia and Yu 2008} \cite{Knight and Fu 2000}   \cite{Zhao and Yu 2006}  \cite{Zou2006}.  However, the conditions for the various penalized likelihood methods to achieve the oracle property are usually not met when the dimension of the feature space has  a polynomial order or an exponential order of the sample size.  The computation also poses a challenge in this case.

 Feature selection procedures of a stepwise nature are computationally appealing.  Efron et al \cite{Efron et al 2004} proposed a sequential procedure called least angle regression (LAR). With slight modification, the algorithm of LAR can also compute the solution path of LASSO sequentially, which made LASSO more popular. The classical forward stepwise regression (FSR) has been recently re-examined in \cite{Wang 2009} on its properties in feature selection with ultra-high dimensional feature space.
A different version of forward stepwise regression referred to as forward selection in \cite{Weisberg1980} has been re-considered recently and dubbed as orthogonal matching pursuit (OMP), see  \cite{Cai and Wang 2010} \cite{Tropp 2004} \cite{Tropp and Gilbert 2007}. The difference between FSR and OMP is that the former selects at each step the covariate that reduces the residual sum of squares the most while the latter selects the covariate that has the largest absolute correlation with the current residuals.

 In many practical problems,  the identification of the set of causal features is of primary interest.  For example, in genetic quantitative trait loci (QTL) mapping and disease gene mapping, of interest are the markers which are either QTL or disease gene themselves or are in linkage disequilibrium with QTL or disease genes.  More relevant properties required of a feature selection method for this purpose are sure screening \cite{Fan and Lv 2008} and  selection consistency.  The sure screening property means that the selected set should contain the set of causal features with probability converging to 1. The selection consistency means that the selected set should be the same as the exact set of causal features with probability converging to 1.  Under the well-known irrepresentable condition, the LASSO has been shown to possess the property of selection consistency while the penalty parameter is properly chosen \cite{Zhao and Yu 2006}.  If  the covariance matrix of the vector of the covariates has  eigenvalues bounded both from above and away from zero in addition to some other assumptions, it is established in \cite{Wang 2009} that the FSR has the sure screening property when the procedure is carried out at a certain step before the number of steps reaches the sample size.  The OMP has been studied under conditions called Exact Recovery Condition (ERC) \cite{Candes and Tao 2007} \cite{Tropp 2004}  and Mutual Incoherence Property (MIP) \cite{Cai and Wang 2010}.  The ERC is similar to the irrepresentable condition but much stronger.  The MIP is the condition that $\rho_{\max} < \frac{1}{2k-1}$ where $\rho_{\max}$ is the largest absolute correlation among all pairs of covariates and $k$ is the number of causal covariates.  The ERC implies MIP, see \cite{Cai and Wang 2010}  \cite{Tropp 2004}.  Both the sure screening property and the selection consistency of OMP have been examined in  \cite{Cai and Wang 2010} under MIP together with other conditions.

We propose an alternative sequential feature selection procedure which we called sequential LASSO.  The procedure starts with the original LASSO and  the penalty parameter is tuned to the largest which  allows some  coefficients to be estimated nonzero.  The features with nonzero estimated coefficients form the current active set.   In the following steps,  a partially penalized sum of squares is considered. The coefficients of the features in the current active set are not penalized but the coefficients of all the other features are.  Then the penalty parameter is again tuned to the largest which allows some features outside the current active set to have  estimated nonzero coefficients.  The current active set is then updated by adding these new features with nonzero  estimated coefficients. The procedure continues until it meets a certain stopping rule. We investigate the properties of the sequential LASSO in this article.  We establish its selection consistency in the situation that the dimension of the feature space is of an exponential order of the sample size and the number of causal features is allowed to diverge under conditions weaker than the conditions mentioned in the last paragraph.  We provide some general special cases where the conditions required for the sequential LASSO to be selection consistent hold but the conditions mentioned in the last paragraph fail.  The sequential LASSO bears some similarity with OMP.  At steps where a partial positive cone condition is satisfied, the sequential LASSO selects new features with the same criterion as OMP.  The properties established for the sequential LASSO then apply to OMP.  Thus, we reveal some new properties of OMP other than those discovered in  \cite{Cai and Wang 2010} \cite{Tropp 2004}   \cite{Tropp and Gilbert 2007} .   The stopping rule is given by the extended BIC (EBIC) proposed in \cite{Chen and Chen 2008}.   The selection consistency of EBIC in the same situation is recently established under similar conditions in \cite{LuoChen2011}. Thus, coupled with EBIC the sequential LASSO provides a practically applicable selection consistent method for feature selection in small-$n$-large-$p$ problems.
Simulation studies are carried out to compare the sequential LASSO with other stepwise methods such as FSR and the original LASSO.

The remainder of the article is arranged as follows.  The detailed procedure of the sequential LASSO and its basic properties are given in \S \ref{sec2}.  The selection consistency of the sequential LASSO is studied in \S \ref{sec3}.   Simulation studies are reported in \S \ref{sec4}.

\section{Procedure of Sequential LASSO and its basic properties}
\label{sec2}

Consider the linear regression model below:
\begin{equation}\label{BasicModel}
  y_i = \beta_0 + \sum_{j=1}^{p_{n}} \beta_{nj} x_{ij} + \epsilon_i,  \ i =1, \dots, n,
\end{equation}
where $\epsilon_i$'s are i.i.d. normal variables with mean zero and variance $\sigma^2$, the $x_{ij}$'s are called features which are either deterministically determined or observed at random.  The following particular natures are assumed for the above model.  (a)  The dimensionality of the feature space is assumed as $\ln p_n = O(n^\kappa)$ for  $\kappa > 0$ (ultra-high).   (b)  Let $s_{0n} = \{ j:  \beta_{nj} \neq 0\}$ and let $|s_{0n}|$ denote the cardinality of $s_{0n}$.  It is assumed that $|s_{0n}| = O(n^c)$ for some $0<c<1$.  (c) The magnitude of $\beta_{nj}, j \in s_{0n},$ is allowed to vary with $n$.    In matrix notation, (\ref{BasicModel}) is expressed as
\[
 \vy_n = X_n \vbeta_n + \vepsilon_n,
 \]
where $\vbeta_n = (\beta_{n1}, \dots, \beta_{np_n})^{\tau}$,  $\vy_n = (y_1, \dots, y_n)^{\tau}$   and $X_n= (x_{ij})_{\stackrel{ i=1,\dots, n} {j=1,\dots,p_n}}$ and $\vepsilon_n = (\epsilon_1, \dots, \epsilon_n)^{\tau}$.   Let the  columns of $X_n$ be normalized such that $\frac{1}{n} \sum_{i=1}^n x_{ij} = 0$ and $\frac{1}{n} \sum_{i=1}^n x_{ij}^2 = n$ for all $j$.  For the sake of simplicity, the subscript $n$ in the notation will be dropped afterwards.
  Let $S$ denote the set of indices $\{1,2,\cdots,p_{n}\}$.  The sequential LASSO is described as follows.
\begin{itemize}
\item The procedure starts with the $L_1$ penalized sum of squares:
\[ l_1 = \| \vy - X \vbeta\|_2^2 + \lambda \sum_{j \in S} |\beta_j|, \]
where $\| \cdot \|_2 $ is the $L_2$ norm of a vector.   $l_1$  is minimized by tuning  $\lambda$ to a value such that it is the largest to allow some $\beta_j$  nonzero in the minimizer. The set of indices of nonzero $\beta_j$'s is denoted by~$s_{*1}$ and referred to as the active set.

\item In the second step, $l_1$ is replaced by
		\[ l_2 = \| \vy - X\vbeta\|_2^2 + \lambda \sum_{j \in s_ {*1}^c} |\beta_j|, \]
where $s_{*1}^c$ denotes the complement of $s_{*1}$ in $S$.  Then $l_2$  is minimized by tuning  $\lambda$ to a value such that it is the largest  to allow some $\beta_j$ with $j  \in s_{*1}^c$ nonzero in the minimizer. The active set is updated as the set of  all features with nonzero coefficient in this minimization and  denoted by $s_ {*2}$.

\item In general, after $k$ steps have been carried out and the active set $s_ {*k}$ is obtained,  the penalized sum of squares
\[l_{k+1} = \| \vy - X \vbeta\|_2^2 + \lambda \sum_{j \in s_ {*k}^c} |\beta_j| \]
is then minimized by tuning  $\lambda$ to a value such that it is the largest  to allow some $\beta_j$ with $j  \in s_ {*k}^c$ nonzero in the minimizer, and the active set is updated as $s_{*k+1}$.

\item The process continues until some stopping criterion is met.
\end{itemize}
The sequential LASSO described above  selects features sequentially by different partially $L_1$ penalized sum of squares.  Once a feature is selected at a certain step, its coefficient will no longer be penalized in the subsequent step, which ensures that the feature will always remain in the model, see the basic properties below.  This differs from the ordinary LASSO where a feature included in an earlier stage could be left out in a later stage in the solution path.

Let  $s$ be any subset of $S$.  Denote by $X(s)$ the matrix consisting of the columns of $X$ with indices in $s$.   Similarly, let $\vbeta(s)$ denote the vector consisting of the corresponding components of $\vbeta$.   Let  ${\cal R}(s)$ be the linear space spanned by the columns of $X(s)$ and
$H(s)$ denote its projection matrix, i.e, $H(s) =X(s)[X^{\tau}(s)X(s)]^{-1}X^{\tau}(s)$.
Some basic properties of the sequential LASSO are given in the following.

\begin{proposition}\label{prop2.1}
For $k\geq 1$ and any $l \in s_{*k}^c$, if $X(\{l\}) \in {\cal R}(s_{*k})$ then $l \not \in s_{*k+1}$.
\end{proposition}
\begin{proof}[Proof:]  If $X(\{l\}) \in {\cal R}(s_{*k})$ then  there exists an $\va_k$ such that $X(\{l\}) = X(s_{*k}) \va_k $ and hence
\begin{eqnarray*}
l_{k+1} & = & \|\vy - X(s_{*k}) (\vbeta(s_{*k}) + \beta_l \va_k) - X(s_{*k}^c/\{l\}) \vbeta( s_{*k}^c/\{l\}) \|_2^2 + \lambda ( |\beta_l|  + \sum_{j \in s_{*k}^c/\{l\}} |\beta|_j )\\
& = &  \|\vy - X(s_{*k}) (\tilde{\vbeta}(s_{*k})  - X(s_{*k}^c/\{l\}) \vbeta( s_{*k}^c/\{l\}) \|_2^2 + \lambda ( |\beta_l|  + \sum_{j \in s_{*k}^c/\{l\}} |\beta|_j) \\
& \leq &  \|\vy - X(s_{*k}) (\tilde{\vbeta}(s_{*k})  - X(s_{*k}^c/\{l\}) \vbeta( s_{*k}^c/\{l\}) \|_2^2 + \lambda \sum_{j \in s_{*k}^c/\{l\}} |\beta|_j.
\end{eqnarray*}
Thus when $l_{k+1}$ is minimized there must be $\beta_l = 0$, i.e., $l \not \in s_{k+1}$.
\end{proof}
\noindent Proposition \ref{prop2.1} implies that, for any $k$, the matrix $X(s_{*k})$ is of full column rank.  It also suggests that, in the sequential LASSO procedure,  any feature that is highly correlated with the features selected already will have little chance to be selected subsequently.
This nature of the sequential LASSO is favorable when it is used for feature selection in ultra-high dimensional feature space where high spurious correlations present, see \cite{Fan and Lv 2008}.

\begin{proposition}
\label{prop2.2}
For $k\geq 1$, the minimization of $l_{k+1}$ is equivalent to the minimization~of
\[ \|[\mathbf{\I}-H (s_{*k})][\vy- X(s_{*k}^c) \vbeta(s_{*k}^c)] \|^2 + \lambda \sum_{j \in s_{*k}^c} |\beta_j|. \]
\end{proposition}

\begin{proof}[Proof:]   Differentiating $l_{k+1}$ with respect to $\vbeta(s_{*k})$, we have
\[
\frac{\partial l_{k+1}} {\partial \vbeta(s_{*k})} = -2X^{\tau}(s_{*k})\vy +2X^{\tau}(s_{*k})X(s_{*k})\vbeta(s_{*k})+2X^{\tau}(s_{*k})X(s_{*k}^c)\vbeta(s_{*k}^c).
\]
Setting the above derivative to zero, we obtain
\begin{equation}\label{eq2.1}
\hat{\vbeta}(s_{*k})=[X^{\tau}(s_{*k})X(s_{*k})]^{-1}X^{\tau}(s_{*k})[\vy - X(s_{*k}^c)\vbeta(s_{*k}^c)].
\end{equation}
Substituting (\ref{eq2.1}) into $\|\vy - X\vbeta\|^2$ we have
\begin{eqnarray*}
l_{k+1} & = & \| \vy - X(s_{*k}) \vbeta(s_{*k}) - X_(s_{*k}^c)\vbeta(s_{*k}^c)\|^2 + \lambda \sum_{j \in s_{*k}^c} |\beta_j|  \\
              & = & \| [\vy \! \!- \!\!X(s_{*k}^c)\vbeta(s_{*k}^c) ]\!\!- \!\!X(s_{*k}) [X^{\tau}\!(s_{*k})X(s_{*k})]^{-1}\!X^{\tau}\!(s_{*k})[\vy\!\! - \!\!X(s_{*k}^c)\vbeta(s_{*k}^c)]\|^2 \!\!+\!\! \lambda \!\sum_{j \in s_{*k}^c} |\beta_j| \\
 & = & \|[\mathbf{\I}-H (s_{*k})][\vy- X(s_{*k}^c) \vbeta(s_{*k}^c)] \|^2 + \lambda \sum_{j \in s_{*k}^c} |\beta_j|.
\end{eqnarray*}
\end{proof}
\noindent As a by-product of the above proof, the components of $\hat{\vbeta}(s_{*k})$ are almost surely nonzero since $\vy$ is a vector of continuous random variables.  This implies that,  in the sequential LASSO, we have $s_{*1} \subset s_{*2} \subset \cdots \subset s_{*k} \subset \cdots$; that is, the models selected in the sequential steps are nested.

 For a general $k$, let $\tilde{\vy} = [I-H(s_{*k})] \vy$, $\tilde{X} = [I-H(s_{*k})] X(s_{*k}^c)$, $\tilde{\vbeta} = \vbeta(s_{*k}^c)$ and $\nu_{\bar{k}} = |s_{*k}^c|$.  Then by Proposition \ref{prop2.2} the minimization of $l_{k+1}$ is equivalent to the minimization of
\begin{equation}
\label{eq2.2}
 \tilde{l}_{k+1} = \|\tilde{\vy} - \tilde{X} \tilde{\vbeta} \|^2 + \lambda \sum_{j =1}^{\nu_{\bar{k}} } |\tilde{\beta}_j|.
\end{equation}
The following proposition is the  Karush-Kuhn-Tucker (KKT) condition for the solution of the above minimization problem.

\begin{proposition}[KKT condition] \label{prop2.3}
Let
\[ \partial |x|  = \left\{  \begin{array}{ll}
                         1, & \mbox{if} \ x >0, \\
                        -1, & \mbox{if} \  x<0, \\
                         r,  & \mbox{if} \  x=0, \end{array} \right.
\]
where $r$ is an arbitrary number with $|r| \leq 1$.
Let $\partial \|\tilde{\vbeta}\|_1 = (\partial |\tilde{\beta}_1|, \dots, \partial |\tilde{\beta}_{\nu_{\bar{k}}}|)^\tau$.  Then $\tilde{\vbeta}$ is a minimizer of (\ref{eq2.2}) if
\[ 2 \tilde{X}^\tau ( \tilde{\vy} - \tilde{X} \tilde{\vbeta}) = \lambda \partial \|\tilde{\vbeta}\|_1.\]
\end{proposition}

\begin{proof}[Proof:] We only need to verify that the form of $\partial \|\tilde{\vbeta}\|_1$ given above is the sufficient and necessary condition for a sub gradient of $\|\tilde{\vbeta}\|_1$.
First, for any $\vxi$,  we have
\begin{eqnarray*}
\|\vxi\|_1-\|\tilde{\vbeta}\|_1 & = & \sum_{j: \xi_j \neq  \tilde{\beta}_{j}}
        (|\xi_j|-|\tilde{\beta}_{j}|)  \\
& \geq  & \sum_{j: \xi_j \neq  \tilde{\beta}_{j} }
     \partial |\tilde{\beta}_j| (\xi_j -\tilde{\beta}_{j})
=  \partial \|\tilde{\vbeta}\|_1^{\tau}(\vxi -\tilde{\vbeta}).
\end{eqnarray*}
Thus by  definition  $\partial \|\tilde{\vbeta}\|_1$ is a sub gradient.

Next, let $\vw$ be any sub gradient of $\|\tilde{\vbeta}\|_1$.  We show that
\[ w_j  = \left\{  \begin{array}{ll}
                         1, & \mbox{if} \ \tilde{\beta}_j >0, \\
                        -1, & \mbox{if} \  \tilde{\beta}_j<0, \\
                         r,  & \mbox{if} \  \tilde{\beta}_j =0.  \end{array} \right.
\]
Suppose $\tilde{\beta}_j =0$ and assume $|w_j| > 1$.  Then we can  define a new
vector $\vxi$ such that
$\xi_j=\mbox{sign}(w_j)$ and $\xi_i = \tilde{\beta}_i$ for $i  \neq j$.   Then we
have
$\|\vxi\|_1-\|\tilde{\vbeta}\|_1=1<\vw^{\tau}(\vxi -\tilde{\vbeta})=|w_j|$,  contradicting to that $\vw$ is a sub gradient.

Now suppose $\tilde{\beta}_j \neq 0$.  For a positive number $\delta < |\tilde{\beta}_j|$, define $\vxi_1$ and $\vxi_2$ such that $\xi_{1j} = \tilde{\beta}_j + \delta \mbox{sign}(\tilde{\beta}_j)$, $\xi_{2j} =  \tilde{\beta}_j - \delta \mbox{sign}(\tilde{\beta}_j)$ and $\xi_{1i}=  \xi_{2i} = \tilde{\beta}_i, i\neq j$.
Since $\vw$ is a sub gradient we must have
\begin{eqnarray*}
& & \|\vxi_1 \|_1-\|\tilde{\vbeta}\|_1 = \delta  \geq \vw^\tau ( \vxi_1 - \tilde{\vbeta}) =  \delta w_j \mbox{sign}(\tilde{\beta}_j ),  \\
& & \|\vxi_2 \|_1-\|\tilde{\vbeta}\|_1 = - \delta \geq \vw^\tau ( \vxi_2 - \tilde{\vbeta}) =  -  \delta w_j \mbox{sign}(\tilde{\beta}_j ),
\end{eqnarray*}
which implies  $w_j \mbox{sign}(\tilde{\beta}_j ) = 1$ and hence
 $w_j=\mbox{sign}(\tilde{\beta}_j).$

\end{proof}

In the remainder of this section, we highlight the difference of the sequential LASSO from FSR and OMP.    First, consider the difference between the sequential LASSO and FSR.  After the sub model $s_{*k}$ is selected, the sequential LASSO  selects the next feature among the features that maximize
\[g _1(j) = |X_j^\tau [ I - H(s_{*k})]\vy|, \]
see the proof of Theorem \ref{theorem1} in \S \ref{sec3}.  The FSR selects the next feature by minimizing $\mbox{RSS}(j) = \vy^\tau [ I - H(s_{*k}\cup \{j\})] \vy$ which is equivalent to maximizing
\[ g_2(j) = \frac{|X_j^\tau [ I - H(s_{*k})]\vy| }{ \sqrt{ X_j^\tau [ I - H(s_{*k})] X_j} }.
\]
The equivalence is established by the following identity
\[ I - H(s_{*k}\cup \{j\}) = [ I - H(s_{*k})]\left( I - \frac{ X_j X_j^\tau[ I - H(s_{*k})] }{  X_j^\tau [I - H(s_{*k}) ]X_j} \right).\]
The sequential LASSO  selects the next feature  that has the highest correlation with the current residual $[ I - H(s_{*k})]\vy$ but the FSR selects the next feature that has the highest inflated correlation with an inflating factor $[ X_j^\tau [ I - H(s_{*k})] X_j]^{-1/2}$.  If $X_j$ is orthogonal to ${\cal R}(s_{*k})$, the factor is a constant (note that the $X_j$'s are standardized), but  larger than the constant otherwise. The more correlated the $X_j$ is with the features in $s_{*k}$,  the larger the  inflating factor. If two features have the same absolute correlation with the current residual, the FSR will select the one that is more correlated with the features in $s_{*k}$.  If one feature has a lower correlation with the current residual but is more correlated with the features in $s_{*k}$ than another feature,  it might turn out that this feature has a higher inflated correlation and is selected by FSR.   Obviously, this is a disadvantage of FSR, especially when high spurious correlations present in small-$n$-large-$p$ problems.

The OMP selects the next feature (or features) maximizing  $g_1(j)$.  At steps where there is only one feature that maximizes $g_1(j)$,  the sequential LASSO and the OMP select the same next feature.  But at steps where there are more than one features that maximize  $g_1(j)$, there is a  difference between the sequential LASSO and the OMP. The OMP selects all those features.   But the sequential LASSO selects them all subject to a partial positive cone condition, see the proof of Theorem \ref{theorem1}.  If the partial positive cone condition is not satisfied, the sequential LASSO generally does not select all those features.  The sequential LASSO can be easily extended as a sequential penalized likelihood method for generalized linear models but there is no obvious way by which the OMP can be extended.  We will explore the properties of extended sequential penalized likelihood method in our future research.

\section{Selection consistency of sequential LASSO with ultra-high dimensional feature space}
\label{sec3}

We establish in this section the selection consistency of the sequential LASSO when the dimension of the feature space is ultra-high, i.e., $\ln p_n = O(n^\kappa), \kappa >0$, under two different settings of the feature matrix $X$: (i) $X$ is deterministic and (ii) $X$ is random.   The deterministic case is dealt with in \S \ref{sec3.1} and  the random case in \S \ref{sec3.2}.  Some interesting special cases are discussed in  \S \ref{sec3.3}.  The sequential LASSO with EBIC as the stopping rule is considered in  \S \ref{sec3.4}

\subsection{The case of deterministic feature matrix}
\label{sec3.1}

In the deterministic case, the columns of $X$ are normalized such that the sample mean and variance of each feature are 0 and $n$ respectively.   We now introduce some notations.  For  $s\subset S$,   let  $s^- = s^c \cap s_{0}$.  Recall that $s_{0}$ is the set of indices of the  nonzero $\beta_j$'s.  If $s \subset s_{0}$ then $s^-$ is the complement of $s$ in $s_{0}$. For $s \subset s_{0}$, define
\[ \gamma_n(j, s, \vbeta) = \frac{1}{n}X_j^\tau [ I - H(s)] X\vbeta. \]
In fact,  $ \gamma_n(j, s, \vbeta)$ only depends on $\vbeta(s^c)$. But for the ease of notation, $\vbeta$ and $\vbeta(s^c)$ will be used interchangeably.
Unless otherwise stated, $\vbeta$ also denotes the unknown true value of the parameter vector. The selection consistency of the sequential LASSO in the case of deterministic feature matrix is established under the following assumptions.
\begin{description}
  \item[A1] $  \max_{j \in s^c_{0}}|\gamma_n(j, s, \vbeta)|<q \max_{j\in s^{-}}|\gamma_n(j, s,\vbeta)|, $ $0 < q < 1$.
\item[A2] (Partial positive cone condition).  Let
 \[ {\cal A}_s= \{\tilde{j} : \tilde{j} \in s^c, |\gamma_n(\tilde{j}, s,\vbeta)|=\max_{j \in
s^c}|\gamma_n(j, s,\vbeta)|\},
\]
and $ \tilde{X}({\cal A}_s)=[I - H(s)]X({\cal A}_s)$. Then
$ [\tilde{X}^{\tau}({\cal A}_s)\tilde{X}({\cal A}_s]^{-1}\vone >0, $
where $\vone$ is the vector with all components 1.
\item[A3]  $\frac{ \sqrt{n} }{\ln p_n} \lambda_{\min}[\frac{1}{n}X^{\tau}(s_{0})X(
s_{0})] \min\limits_{j \in
s_{0}}|\beta_{j}| \rightarrow
+\infty,  \ \text{as}\ n \rightarrow  \infty$, where $\lambda_{\min}$ denotes the smallest eigenvalue.
\end{description}

Assumption A1 is implied by the following condition
\begin{equation}
\label{eq3.1}
 \|\tilde{X}^{\tau}_j\tilde{X}(s^{-})[\tilde{X}^{\tau}(s^{-})\tilde{X}(s^{-})]^{-1}\|_1<1-\eta ,\forall j\in
s_{0}^c,
\end{equation}
where $\tilde{X}_j = [I - H(s)]X_j$ and $0< \eta < 1$.   The claim above follows because
\begin{eqnarray*}
|\gamma_n (j, s,\vbeta)| & =& \frac{1}{n} |X_j^{\tau}[I-H(s)]{\vmu}|   \\
& =& |\tilde{X}^{\tau}_j \tilde{X}(s^{-})[\tilde{X}^{\tau}(s^{-})\tilde{X}(s^{-})]^{-1}\frac{1}{n}\tilde{X}^{\tau}(s^{-})[I - H(s)]\vmu | \\
& \leq & \| \tilde{X}^{\tau}_j \tilde{X}(s^{-})[\tilde{X}^{\tau}(s^{-})\tilde{X}(s^{-})]^{-1}\|_1\frac{1}{n} \| \tilde{X}^{\tau}(s^{-})[I - H(s)]\vmu \|_{\infty} \\
& < & (1-\eta) \frac{1}{n} \|\tilde{X}^{\tau}(s^{-})[I - H(s)]\vmu\|_{\infty}= (1-\eta) \frac{1}{n}  \max_{j\in
s^{-}}|X_j^{\tau}[I-H(s)]\vmu|  \\
&=& (1-\eta) \max_{j\in s^{-} }|\gamma_n(j, s, \vbeta)|,
\end{eqnarray*}
where the strict inequality holds  by (\ref{eq3.1}).

Under assumption A1, the ${\cal A}_s$ in A2 is a subset of $s_0$.  Assumption A2 holds if and only if
\begin{equation}
\label{eq3.2}
\tilde{X}^{\tau}_j\tilde{X}({\cal A}_s\backslash\{j\})[\tilde{X}^{\tau}({\cal A}_s\backslash\{j\})\tilde{X}({\cal A}_s\backslash\{j\})]^{-1}\vone <1,\forall  j\in \mathcal{A}_s.
\end{equation}
  We establish the equivalence of A2 and (\ref{eq3.2}) below.  Let $A = \tilde{X}({\cal A}_s\backslash\{j\})$ and $\vb = \tilde{X}_j$.  Since a permutation of the rows and columns does not change the sum of the rows,  it suffices to verify that the sum of the last row of $\begin{pmatrix}
A^{\tau}A & A^{\tau}\vb\\
\vb^{\tau}A & \vb^{\tau}\vb
\end{pmatrix}^{-1}$ is positive if and only if   $\vb^{\tau} A (A^\tau A)^{-1} \vone <1$.  Let $E= I - A(A^\tau A)^{-1} A^\tau$ and $F=  I - \vb(\vb^\tau \vb)^{-1} \vb^\tau$.   By the formula for the inverse of blocked matrices, we have
\[ \begin{pmatrix}
A^{\tau}A & A^{\tau}\vb\\
\vb^{\tau}A & \vb^{\tau}\vb
\end{pmatrix}^{-1} =
\begin{pmatrix}
(A^{\tau}FA)^{-1} & -(A^{\tau}A)^{-1}A^{\tau}\vb(\vb^{\tau}E\vb)^{-1}\\
-(\vb^{\tau}\vb)^{-1}\vb^{\tau}A(A^{\tau}FA)^{-1} &
(\vb^{\tau}E\vb)^{-1}
\end{pmatrix}.
\]
and
\begin{eqnarray*}
   (A^{\tau}FA)^{-1}   & = & [A^\tau A - A^\tau\vb (\vb^\tau \vb)^{-1} \vb^\tau A]^{-1} \\
     & = & (A^\tau A)^{-1} +  (A^\tau A)^{-1} A^\tau ( \vb^\tau E \vb)^{-1}\vb^{\tau}A   (A^\tau A)^{-1} .
     \end{eqnarray*}
Substituting the expression of $   (A^{\tau}FA)^{-1}   $ into the first block of the last row of the above matrix, we obtain
\[ -(\vb^{\tau}\vb)^{-1}\vb^{\tau}A(A^{\tau}FA)^{-1}  = -( \vb^{\tau}E\vb)^{-1}\vb^{\tau} A ( A^\tau A)^{-1}. \]
Thus the sum of the last row becomes
\[ (\vb^{\tau}E\vb)^{-1} - ( \vb^{\tau}E\vb)^{-1}\vb^{\tau} A ( A^\tau A)^{-1} \vone =  (\vb^{\tau}E\vb)^{-1}[1 - \vb^{\tau} A ( A^\tau A)^{-1} \vone] \]
which is greater than 0 if and only if $ \vb^{\tau} A ( A^\tau A)^{-1} \vone < 1$.

Condition (\ref{eq3.1}) is a conditional version of ERC conditioning on the subset $s$ of the relevant features.    Condition (\ref{eq3.2}) is similar to but much weaker than the {\it irrepresentable condition}.  The above arguments suggest that Conditions A1 and A2 might be weaker than the ERC and the {\it irrepresentable condition}.  This is indeed the case. We will demonstrate this by special cases in \S \ref{sec3.3} where the conditions for the selection consistency of the sequential LASSO hold but the ERC and the {\it irrepresentable condition} are not satisfied. If $\lambda_{\min}(\frac{1}{n}X^{\tau}(s_{0})X^{\tau}( s_{0}))$ is bounded   away from zero,  which is a common assumption in the case of ultra-high dimensional feature space, then Condition A3 is equivalent to
$\frac{\sqrt{n}}{\ln p_n} \min\limits_{j\in
s_{0}}|{\beta}_j|\rightarrow \infty$. If $\ln p_n = O(n^\kappa)$ with $\kappa < 1/2$ and  $\min\limits_{j\in
s_{0}}|{\beta}_j| \geq C n^{-\delta}$ for some constant $C$ and $\delta < 1/2 - \kappa$, A3 is then satisfied.

\vspace{0.2in}
We now state and prove the major theorem in the following.

\begin{theorem}
\label{theorem1}
Suppose that assumptions A1-A3 hold. Let $\ln p_n = O(n^\kappa)$, where $\kappa < 1/2$. Then the sequential LASSO is selection consistent in the sense that
\[ Pr (s_{*k^*} = s_0)  \to 1,  \ \  \mbox{as} \ \ n \to \infty, \]
where $s_{*k^*}$ is the set of features selected at the $k^*$th step of the sequential LASSO such that $|s_{*k^*}| = p_0$, $s_0$ is the set of relevant features and
 $p_0 = |s_0|$.
\end{theorem}

\begin{proof}[Proof]  By Proposition \ref{prop2.3}, at the $(k+1)$st step of the sequential LASSO,  the solution $\hat{\vbeta}$  satisfies
\begin{equation}\label{eq3.3}
 2 \tilde{X}^\tau ( \tilde{\vy} - \tilde{X} \hat{\vbeta}) = \lambda \partial \|\hat{\vbeta}\|_1,
\end{equation}
where $\tilde{\vy} = [I-H(s_{*k})] \vy$, $\tilde{X} = [I-H(s_{*k})] X(s_{*k}^c)$, and $\partial \|\hat{\vbeta}\|_1$ is a sub gradient of $\|\vbeta\|_1$ at $\hat{\vbeta}$ whose components are $1, -1$ or a number with absolute value less than or equal to 1 according as the components are positive, negative or zero.  For $k=0$, $s_{*0}$ is taken as the empty set $\phi$. Obviously, $s_{*0} \subset s_0$.  Assume that $s_{*k} \subset s_0$ and $|s_{*k}| < p_0$.
Let
\[  \hat{\gamma_n}(j, s_{*k}, \vbeta) = \frac{1}{n} X_j^\tau [I - H(s_{*k})] \vy  = \gamma_n(j, s_{*k}, \vbeta) + \frac{1}{n}X_j^\tau [I - H(s_{*k})] \epsilon.
\]
Define
\[ {\cal A}_k=\{j:|\hat{\gamma_n}(j,s_{*k},\vbeta)|=\max_{j\in
s_{*k}^c}|\hat{\gamma_n}(j, s_{*k},\vbeta)|\}.
\]
We are going to show that, with probability converging to 1,  ${\cal A}_k \subset s_0$ and that ${\cal A}_k$ is the set of non-zero elements of the solution  to equation (\ref{eq3.3}).
We first show
 that ${\cal A}_k \subset s_0$, which is implied by    $|\hat{\gamma_n}(j, s_{*k}, \vbeta)| >  \max_{l \in
s_0^c }|\hat{\gamma_n}(l, s_{*k},\vbeta)| $  for $j \in s_{*k}^-$ with probability converging to 1.   The statement is established by showing
\begin{description}
  \item[(i)] $\frac{1}{n}X_j^\tau [I - H(s_{*k})] \epsilon = O_p(n^{-1/2}\ln p_n)$ uniformly for all $j \in s_{*k}^c$.
  \item[(ii)] For $j  \in s_{*k}^-$,   $\max_{j\in
s_{*k}^-}|\gamma_n(j, s_{*k},\vbeta)| \geq C_n n^{-1/2}\ln p_n$ for $C_n \to \infty$.
\end{description}
Notice that $X_j^\tau [I - H(s_{*k})] \epsilon \sim N(0, \sigma^2 \|\tilde{X}_j\|_2^2) $ where $\|\tilde{X}_j\|_2^2 \leq \|{X}_j\|_2^2 = n$.   Hence
\begin{eqnarray*}
& & P( \frac{1}{n} |X_j^\tau [I - H(s_{*k})] \epsilon | > \sigma n^{-1/2}\ln p_n ) \\
  &= & P( |X_j^\tau [I - H(s_{*k})] \epsilon | > \sigma n^{1/2}\ln p_n )  \\
& \leq  & P( |X_j^\tau [I - H(s_{*k})] \epsilon | > \sigma \|\tilde{X}_j\|_2 \ln p_n )  \\
& = & P( |z | >  \ln p_n )  \leq  \frac{2}{\ln p_n} \exp \{ - \frac{(\ln p_n)^2}{2} \},
\end{eqnarray*}
where $z$ is a standard normal random variable. Thus,  by Bonferroni inequality,
\begin{equation}
\label{eqT1}
P( \max_{j \in s_{*k}^c} \frac{1}{n}|X_j^\tau [I - H(s_{*k})] \epsilon | > \sigma n^{-1/2}\ln p_n )
 \leq  \frac{2}{\ln p_n} \exp \{ - \frac{(\ln p_n)^2}{2} + \ln p_n \} \to 0.
\end{equation}
Thus (i) is proved.

Let $\Delta (s_{*k}) = \vmu^\tau [I - H(s_{*k})] \vmu$ where $\vmu = X \vbeta$.  We have the following inequalities
\begin{equation}
\label{eq3.4}
\Delta(s_{*k}) =\sum_{j \in s_{*k}^{-} }\beta_{j} X_j^{\tau}[I-H(s_{*k})] \vmu
\leq  \|\vbeta(s_{*k}^-) \|_1 \max_{j \in
s_{*k}^- }  |\gamma_n(j, s_{*k},\vbeta)|,
 \end{equation}
and
\begin{equation}
\label{eq3.5}
\begin{split}
\Delta(s_{*k})=&\vbeta^\tau(s_{*k}^-) X^{\tau}(s_{*k}^-)[I-H(s_{*k})]X(s_{*k}^-)\vbeta (s_{*k}^-)\\
\geq &
\lambda_{\min}( X^{\tau}(s_{*k}^-)[I-H(s_{*k})]X(s_{*k}^-) ) \| \vbeta (s_{*k}^-)\|_2^2\\
\geq &\lambda_{\min}(X^{\tau}(s_{0})X(
s_{0}))\| \vbeta (s_{*k}^-)\|_2^2.
\end{split}
\end{equation}
The second inequality above follows since $ s_{*k} \cup s_{*k}^- = s_0$ and
$(X^{\tau}(s_{*k}^-)[I-H(s_{*k})]X(s_{*k}^-) )^{-1}$
is a sub-matrix of $(X^{\tau}(s_{0})X(
s_{0})^{-1}$ by the formula of the inverse of blocked matrices.  Combining (\ref{eq3.4}) and (\ref{eq3.5}) yields
\begin{eqnarray*}
  \max_{j \in s_{*k}^- }  |\gamma_n(j, s_{*k},\vbeta)| & \geq & \lambda_{\min}(\frac{1}{n} X^{\tau}(s_{0})X(
s_{0})) \frac{\| \vbeta (s_{*k}^-)\|_2^2}{ \|\vbeta(s_{*k}^-) \|_1}  \\
& \geq & \lambda_{\min}(\frac{1}{n}X^{\tau}(s_{0})X(
s_{0})) \min_{j \in s_0} |\beta_{j}| \\
& \equiv & C_n n^{-1/2} \ln p_n, \ \ \mbox{say,}
\end{eqnarray*}
with $C_n = \frac{ n^{1/2}}{ \ln p_n} \lambda_{\min}(\frac{1}{n}X^{\tau}(s_{0})X(
s_{0})) \min_{j \in s_0} |\beta_{j}|$.  The second inequality above holds  since $|s_{*k}^-| \| \vbeta (s_{*k}^-)\|_2^2  \geq  \| \vbeta (s_{*k}^-)\|_1^2 \geq |s_{*k}^-| \min_{j \in s_0} |\beta_{0j}| \| \vbeta (s_{*k}^-)\|_1.$  $C_n \to \infty$ by A3. Thus (ii) is proved.

By A1 and (ii),
\begin{eqnarray*}
& & |\max_{j \in s_{*k}^- }  |\gamma_n(j, s_{*k},\vbeta)| - \max_{j \in s_0^c}  |\gamma_n(j, s_{*k},\vbeta)| |  \\
& > & (1-q) \max_{j \in s_{*k}^- }  |\gamma_n(j, s_{*k},\vbeta)|  \geq (1-q) C_n n^{-1/2} \ln p_n.
\end{eqnarray*}
This fact and (i) then imply that  $\hat{\gamma_n}(j,s_{*k},\vbeta)$ must attain the maximum within $s_{*k}^-$. Therefore, ${\cal A}_k \subset s_{*k}^- \subset s_0$.

 Without loss of generality, assume that
$\hat{\gamma_n}(j, s_{*k}, \vbeta) >0$ for
all $j\in {\cal A}_k.$  Consider $\hat{\gamma_n} (j, s_{*k}, \vxi)$ as a function of $\vxi$.  Since the function is continuous, for each $j \in {\cal A}_k$,  there exist a neighborhood   ${\cal N}_j = \{\vxi: \|\vxi -\vbeta\|_2 \leq \delta_j\}$ and a constant $c_j>0 $  such that, for all $\vxi \in {\cal N}_j$,
$\hat{\gamma_n}(j,s_{*k}, \vxi) -\max_{ l \in {\cal A}_k^c}|\hat{\gamma_n}(l, s_{*k}, \vxi))|>c_j$.  Here ${\cal A}_k^c$ denotes the complement of ${\cal A}_k $ in $s_{*k}^c$ by an abuse of notation.
Let ${\cal N} =   \{\vxi:  \|\vxi -\vbeta\|_2 \leq \delta \}$ where $\delta = \min \delta_j$. Then for all $\vxi \in {\cal N}$, $\min_{ j \in {\cal A}_k}\hat{\gamma_n}(j,s_{*k}, \vxi) -\max_{ l \in {\cal A}_k^c}|\hat{\gamma_n}(l, s_{*k}, \vxi))|>C$,
where $ C = \max c_j$.

Now construct $\hat{\vbeta}$ as follows. Let $\hat{\vbeta}({\cal A}_k) = \omega [ \tilde{X}^\tau ({\cal A}_k)    \tilde{X}({\cal A}_k)  ]^{-1} \vone $ and $\hat{\vbeta}({\cal A}_k^c) = 0$, where $\omega >0$.  By A2, $\hat{\vbeta}({\cal A}_k)>0$.  Take $\omega$ small enough such that  $\vbeta - \hat{\vbeta} \in {\cal N}$.
Thus we have $\min_{ j \in {\cal A}_k}\hat{\gamma_n}(j,s_{*k}, \vbeta - \hat{\vbeta}) > \max_{ l \in {\cal A}_k^c}|\hat{\gamma_n}(l, s_{*k}, \vbeta - \hat{\vbeta}))|$.  On the other hand,  for any $j \in {\cal A}_k$,
\begin{eqnarray*}
 \hat{\gamma_n}(j,s_{*k}, \vbeta - \hat{\vbeta})   & = & \max_{j \in s_{*k}^c}\hat{\gamma_n}(j,s_{*k}, \vbeta) - \omega \frac{1}{n} \tilde{X}_j^\tau  \tilde{X}({\cal A}_k) [ \tilde{X}^\tau ({\cal A}_k)    \tilde{X}({\cal A}_k)  ]^{-1} \vone \\
 & = & \max_{j \in s_{*k}^c} \hat{\gamma_n}(j,s_{*k}, \vbeta) - \frac{\omega}{n}.
\end{eqnarray*}
Let  $\lambda= 2 n [ \max_{j \in s_{*k}^c} \hat{\gamma_n}(j,s_{*k}, \vbeta) - \frac{\omega}{n}]$. Then, we have
\begin{eqnarray*}
\label{equation12}
& & 2\tilde{X}_j^{\tau} (\tilde{\vy} - \tilde{X} \hat{\vbeta}) = \lambda,   \ \ \mbox{for} \ \  j\in  {\cal A}_k, \\
& & 2\tilde{X}_j^{\tau} (\tilde{\vy} - \tilde{X} \hat{\vbeta}) < \lambda,   \ \ \mbox{for} \ \  j \not\in  {\cal A}_k.
\end{eqnarray*}
Let $\partial |\hat{\beta}_j| = 2\tilde{X}_j^{\tau} (\tilde{\vy} - \tilde{X} \hat{\vbeta}) / \lambda$ for $j \not \in {\cal A}_k$, and 1  for $j  \in {\cal A}_k$. Then $\partial \|\hat{\vbeta}\|_1$ with these components is a sub gradient of $\|\vbeta\|_1$ at $\hat{\vbeta}$ and  $\hat{\vbeta}$ solves equation (\ref{eq3.3}).  From the construction of $\hat{\vbeta}$, all the features corresponding to the non-zero components of $\hat{\vbeta}$ belong to $s_0$.  Hence $s_{*k+1} \subset s_0$.  Thus we have shown that,  given $s_{*k} \subset s_0$,  $s_{*k+1} \subset s_0$ with probability converging to 1.

If $p_0$ is bounded then we have already established the selection consistency of the sequential LASSO.  If $p_0$ diverges as $n \to \infty$, we need to show that $s_{*k} \subset s_0$, $k=1, \dots, p_0$, simultaneously, with probability converging to 1.  Note that, under the assumptions,   $s_{*k+1} \subset s_0$ is equivalent to  $\min_{ j \in {\cal A}_k}\hat{\gamma_n}(j,s_{*k}, \vbeta) > \max_{ l \in {\cal A}_k^c}|\hat{\gamma_n}(l, s_{*k}, \vbeta))|$ which is implied by
$P( \max_{j \in s_{*k}^c} \frac{1}{n}|X_j^\tau [I - H(s_{*k})] \epsilon | > \sigma n^{-1/2}\ln p_n )   \to 0$.
Therefore, when $p_0$ is divergent, the selection consistency is established if
\[ P( \max_{0 \leq k < p_0} \max_{j \in s_{*k}^c} \frac{1}{n} |X_j^\tau [I - H(s_{*k})] \epsilon | > \sigma n^{-1/2}\ln p_n  ) \to 0,  \ \ \mbox{as} \ n \to \infty.
\]
It follows from (\ref{eqT1}) and the Bonferroni inequality that
\begin{eqnarray*}
& &  P( \max_{0 \leq k < p_0} \max_{j \in s_{*k}^c} \frac{1}{n}|X_j^\tau [I - H(s_{*k})] \epsilon | > \sigma n^{-1/2}\ln p_n  ) \\
& \leq  & \frac{2p_0}{\ln p_n} \exp \{ - \frac{(\ln p_n)^2}{2} + \ln p_n \} \\
& \leq  & \frac{2}{\ln p_n} \exp \{ - \frac{(\ln p_n)^2}{2} + 2 \ln p_n \} \to 0,
\end{eqnarray*}
since $p_0  < p_n$. The proof is completed.
\end{proof}

\subsection{The case of random feature matrix}
\label{sec3.2}

Instead of considering $X$ as a fixed design matrix, we now assume  $\vx_i = (x_{i1}, \dots, x_{ip_n})^\tau$, $i=1, \dots, n$,  are i.i.d. copies of a random vector $\vz  = (z_1, \dots, z_{p_n})^\tau$.  Without loss of generality, assume that $E \vz = 0$ and $\var(\vz) = \Sigma$ with diagonal elements 1 and off-diagonal elements independent of $n$.   Assume that
\begin{description}
  \item[a1]  The off-diagonal elements of $\Sigma$ are bounded by a constant less than 1; that is, the correlation between any two features are bounded below from $-1$ and above from~1.
  \item[a2]  $\sigma_{\max}\equiv \max_{1 \leq j,k\leq p_n} \sigma (z_jz_k)  < \infty$ where $\sigma(z_jz_k)$ denotes the standard deviation of $z_jz_k$.
  \item[a3]  $\max_{1\leq j,k\leq p_n}E \exp(t
z_jz_k) $ and  $\max_{1\leq j\leq p_n}E
\exp(tz_j\epsilon) $ are finite   for $t$ in a neighborhood of zero.
\end{description}
For any $s, \tilde{s} \subset  S$, denote by $\Sigma_{s\tilde{s}}$ the sub matrix of $\Sigma$ with row indices in $s$ and column indices in $\tilde{s}$. Define
\[ \Gamma(j, s, \vbeta) = ( \Sigma_{j S} - \Sigma_{js}\Sigma_{ss}^{-1} \Sigma_{sS}) \vbeta. \]
The following assumptions are imposed:
 \begin{description}
   \item[A$1^{'}$]  For any $s \subset s_0$, $s \neq s_0$,
$ \max_{j\in
s^c_{0}}|\Gamma(j, s,\vbeta)|<\max_{j\in
s^{-}}|\Gamma(j, s,\vbeta)|.
$
\item[A$2^{'}$] Let
$ {\cal A}_s=\{ j: j \in s^c, |\Gamma (j, s,\vbeta)|=\max_{l \in
s^c}|\Gamma(l, s,\vbeta)|\}.
$
Then
 \[ (\Sigma_{{\cal A}_s {\cal A}_s}-\Sigma_{{\cal A}_s s}\Sigma_{s s}^{-1}\Sigma_{s {\cal A}_s})^{-1}\vone>0.
\]
\item[A$3^{'}$] $
 \frac{n^{1/2}}{\ln p_n} \lambda_{\min}(\Sigma_{s_{0}s_{0}})(\min_{j\in
s_{0}}|\beta_j|) \rightarrow
+\infty\;\text{as}\;n\rightarrow +\infty.
$
\end{description}
The assumptions A$1^{'}$ - A$3^{'}$ are in fact the assumptions A1-A3 with the empirical variances and covariances of the features replaced by their theoretical counterparts.  In order to establish the selection consistency of the sequential LASSO in the case of random feature matrix, we need to pass from assumptions A$1^{'}$ - A$3^{'}$ to assumptions A1-A3.  The following lemma  ensures that if A$1^{'}$ - A$3^{'}$ hold then A1-A3 hold with probability converging to 1 as $n$ goes to infinity.

\begin{lemma}
\label{lemma3.2.1}
 Under assumptions $a1$-$a3$,
\begin{description}
  \item[(i) ]  $P(\max_{1\leq j,k\leq p_n
}\left| \frac{1}{n} \sum_{i=1}^n x_{ij}x_{ik}-\Sigma_{jk}\right| >
n^{-\frac{1}{3}} \sigma_{\max})\rightarrow 0.$ \\
\item[(ii)] $ P(\max_{1\leq j\leq p_n}\left|\frac{1}{n} \sum_{i=1}^nx_{ij}\epsilon_i\right| >
n^{-\frac{1}{3}} \sigma)\rightarrow 0.$
\item[(iii)] Let $\Sigma_{jl|s}=\Sigma_{jl}-\Sigma_{js}\Sigma_{ss}^{-1}\Sigma_{sl} $  and $ \hat{\Sigma}_{jl|s} =X_{j}^{\tau}[I-H(s) ]X_{l}/n$. Then
\[
\max_{1\leq j, l \leq p_{n}}\max_{s: |s| \leq
p_{0}}| \hat{\Sigma}_{jl|s} -\Sigma_{jl|s}|=o_p(1).
\]
\end{description}
\end{lemma}

\begin{proof}[Proof]:  For any $j, k\in \{1,2,\cdots, p_n\}$ it follows from \cite{Fill 1983} that
\begin{equation}
\label{eq3.3.1}
P(|\sum_{i=1}^n x_{ij}x_{ik}-n\Sigma_{jk}|> \sqrt{n} \sigma(z_jz_k) \psi_n)\leq
C[1-\Phi(\psi_n)] \exp[\dfrac{\psi_n^3}{\sqrt{n}}\lambda(\dfrac{\psi_n}{\sqrt{n}})]
\end{equation}
where $C$ is a constant, $\Phi(\cdot)$ is the cumulative distribution function of standard normal distribution,
$\lambda(\cdot)$ is the Cramer series for the distribution of $z_jz_k$ which converges in a neighborhood of zero under assumption $a3$, and $\psi_n$ is a sequence satisfying $ \psi_n=o(n^{1/2})$ and $\psi_n \rightarrow  \infty. $

Now take $\psi_n = n^{\frac{1}{6} - \delta}$ for   $0< \delta  < \frac{1}{6} - \frac{\kappa}{2}$. Then
$\lambda(\dfrac{\psi_n}{\sqrt{n}})$ is bounded and
$\dfrac{\psi_n^3}{\sqrt{n}}$ goes to $0$ as $n$ converges to
$\infty$. Thus (\ref{eq3.3.1}) leads to
\begin{eqnarray*}
& & P(|\sum_{i=1}^n x_{ij}x_{ik}-n\Sigma_{jk}|> n^{\frac{2}{3}-\delta} \sigma_{\max} ) \\
 & \leq & P(|\sum_{i=1}^n x_{ij}x_{ik}-n\Sigma_{jk}|> n^{\frac{2}{3}-\delta} \sigma(z_jz_k) ) \\
& \leq & C_1 [1-\Phi(n^{\frac{1}{6} - \delta} )]  \\
& \leq & \frac{C_1}{ n^{\frac{1}{6} - \delta}} \exp ( - \frac{1}{2} n^{\frac{1}{3} - 2\delta}),
\end{eqnarray*}
where $C_1$ is a generic constant.  Let $p_n = \exp( a n^\kappa)$ where $a>0$ and $\kappa < \frac{1}{3}$.
By Bonferroni inequality,
\[
P(\max_{1\leq j,k\leq p_n
}\left|  \sum_{i=1}^n x_{ij}x_{ik}-n\Sigma_{jk}\right| >
 n^{\frac{2}{3}-\delta} \sigma_{\max})  = o( n^{-\frac{1}{6} +\delta }) \to 0.
\]
Hence (i) is proved. The proof of (ii) is similar and is omitted.

Note that, for  $X_j$, $X_l$ and $X(s)$, $\frac{1}{n} X_j^\tau( I - X(s)[X^\tau(s)X(s)]^{-1}X^\tau(s))X_l$ is a continuous function of the means $\frac{1}{n}\sum_{i=1}^n x_{ij} x_{il}$, $\frac{1}{n}\sum_{i=1}^n x_{ij} x_{ik}$, $\frac{1}{n}\sum_{i=1}^n x_{il} x_{ik}$ and $\frac{1}{n}\sum_{i=1}^n x_{ik} x_{im}$,  $k, m \in s$.  Let $\bar{X}_{jls}$ denote the vector consisting of these means and $\vmu_{jls}$ its expectation. The function depends on $|s|$ but not on $n$.  Let $g_{|s|}(\bar{X}_{jls})$ denote this function. We then have $g_{|s|}(\vmu_{jls}) = \Sigma_{jl|s}$.

By  assumption $a1$, the range of  $\vmu_{jls}$ for all $j,l,s$ with fixed $|s|$ is compact. Hence $g_{|s|}$ is also uniformly continuous for all $(j,l,s)$ with fixed $|s|$.  Thus for any $\eta > 0$ there is a $\zeta >0$ such that if $\|\bar{X}_{jls} - \vmu_{jls}\|_{\infty} \leq \zeta$  then $ | g_{|s|}( \bar{X}_{jls}) - g_{|s|}( \vmu_{jls}) | \leq \eta$, where $\zeta$ does not depend on $(j,l,s)$. From the proof of (i), we can choose a $n_0$ such that when $n > n_0$,
\[ P(\max_{1\leq j,k\leq p_n
}\left|  \frac{1}{n} \sum_{i=1}^n x_{ij}x_{ik}-\Sigma_{jk}\right| > \zeta
 )  = o( n^{-\frac{1}{6} +\delta }).
\]
Thus we have
\[ P ( \max_{j,l} | g_{|s|}( \bar{X}_{jls}) - g_{|s|}( \vmu_{jls}) | >  \eta ) = o( n^{-\frac{1}{6} +\delta }).\]
By Bonferroni inequality,
\[ P ( \max_{j,l} \max_{s: |s| \leq p_0 } | g_{|s|}( \bar{X}_{jls}) - g_{|s|}( \vmu_{jls}) | >  \eta )  \leq   o( n^{-\frac{1}{6} +\delta }) p_0\to 0,\]
for $p_0 = O(n^{\frac{1}{6} - \delta})$.  (iii) is proved.

\end{proof}

\begin{theorem}
Let $\ln p_n = O(n^\kappa)$, $\kappa < 1/3$, and $p_0 = O(n^c)$, $\kappa/2 < c < 1/6$.   The sequential LASSO is selection consistent with random feature matrices that satisfy conditions $a1$-$a3$ and A$1^{'}$-A$3^{'}$.
\end{theorem}

The theorem is in fact a corollary of Lemma \ref{lemma3.2.1}.  It follows from the lemma immediately that if $a1$-$a3$ and  A$1^{'}$-A$3^{'}$ are satisfied then A1-A3 hold with probability converging to 1.   Thus the selection consistency of the sequential LASSO with random feature matrix is established.

\subsection{Special cases}
\label{sec3.3}

In this sub section, we provide two special cases where the conditions for the selection consistency of the sequential LASSO can be directly verified. The first special case concerns constant positive correlation among the features. In this case, for the {\it irrepresentable condition} to be satisfied,  some restriction must be imposed.  But such restriction is not needed for sequential LASSO. The second special case deals with a correlation structure under which the {\it irrepresentable condition} is violated.

\vspace{0.2in}
\noindent{\bf Special case I}:  Let the correlation matrix of $\vz$ be given by
\[ \Sigma = (1-\rho)I + \rho \vone \vone^\tau, \]
where $I$ is the identity matrix of  dimension $p_n$,  $\vone$ is a $p_n$-vector of all elements 1, and $0 < \rho \leq \rho_0 < 1$. Note that  $\rho$ is allowed to depend on $n$. But for the ease of notation we don't make this dependence explicit.  In this case,
the assumptions A$1^{'}$-A$3^{'}$ are satisfied with $\min_{j \in s_0} |\beta_j| = Cn^{-1/2+\delta}$ for some constant C and an arbitrarily small positive $\delta$.  The claim is verified in the following.

For any $s \subset S$, the sub correlation matrix $\Sigma_{ss}$ has eigenvalues $1-\rho$ and $1+(|s|-1)\rho$ with multiplicities  $|s|-1$ and 1 respectively. The eigenvector corresponding to $1+(|s|-1)\rho$ is $\vone$ with dimension $|s|$.
The smallest eigenvalue is $1-\rho$.  Thus A$3^{'}$ follows immediately.

Now suppose $s \subset s_0$.  For  any $j, k \in s^c,$ we have
\begin{eqnarray*}
\Sigma_{jk} - \Sigma_{js}\Sigma_{ss}^{-1}\Sigma_{sk}
& = & \Sigma_{jk} - \rho^2\vone^\tau \Sigma_{ss}^{-1} \vone =\Sigma_{jk} - \frac{\rho^2|s|}{1+(|s|-1)\rho} \\
& = & \left\{ \begin{array}{ll}
             \dfrac{(1-\rho)(\rho |s|+1)}{1+(|s|-1)\rho}\equiv a,  & \mbox{if} \  j=k\\
             \dfrac{\rho(1-\rho)}{1+(|s|-1)\rho}\equiv b,  & \mbox{if} \ j\neq k.
\end{array}\right.
\end{eqnarray*}
 Therefore,
\begin{eqnarray*}
\gamma_n(j, s, \vbeta) & = & \sum_{k\in
s^{-}} \beta_k(\Sigma_{jk}-\Sigma_{js}\Sigma_{ss}^{-1}\Sigma_{sk}) \\
&= & \left\{
\begin{array}{ll}
  (a -b) \beta_j + b \sum_{k\in s^{-}}\beta_k  =b  \sum_{k\in s^{-}}\beta_k
+(1-\rho) \beta_j, & \mbox{for} \ j\in s^{-}, \\
b \sum_{k \in s^{-}}\beta_k,  &  \mbox{for} \ j\in s_{0}^c.
\end{array} \right.
\end{eqnarray*}
Thus
\[ \max_{j \in s^{-}} |\gamma_n(j, s, \vbeta)|= \left\{ \begin{array}{ll}
              |b  \sum_{k\in s^{-}}\beta_k|
+(1-\rho) \max_{j \in s^{-}} \beta_j & \mbox{if } \ \sum_{k\in s^{-}}\beta_k>0, \\
 |b  \sum_{k\in s^{-}}\beta_k|
+(1-\rho) |\min_{j \in s^{-}} \beta_j| & \mbox{if } \ \sum_{k\in s^{-}}\beta_k<0.
\end{array} \right.
\]
Obviously, $\max_{j \in s^{-}} |\gamma_n(j, s, \vbeta)| > \max_{j \in s_0^c}  |\gamma_n(j, s, \vbeta)|$ and hence A$1^{'}$ is satisfied.
Finally, we have
\begin{eqnarray*}
& &  \Sigma_{{\cal A}_s {\cal A}_s } - \Sigma_{{\cal A}_s s} \Sigma_{ss}^{-1} \Sigma_{s{\cal A}_s }  \\
& = & (1-\rho)I + \rho \vone \vone^\tau - \rho^2 \vone \vone^\tau \Sigma_{ss}^{-1}\vone \vone^\tau \\
& = & (1-\rho)I + \rho \vone \vone^\tau - \frac{\rho^2|s|}{1+(|s|-1)\rho} \vone \vone^\tau \\
& = & (1-\rho)I +  \frac{\rho(1-\rho)|}{1+(|s|-1)\rho} \vone \vone^\tau.
\end{eqnarray*}
Let $\nu$ be the number of elements in ${\cal A}_s$. The eigenvalue of the above matrix corresponding to the eigenvector $\vone$ is
\[ 1- \rho + \frac{\nu \rho (1-\rho)}{ 1+(|s|-1)\rho} = a + (\nu -1 ) b.\]
Hence
\[ ( \Sigma_{{\cal A}_s {\cal A}_s } - \Sigma_{{\cal A}_s s} \Sigma_{ss}^{-1} \Sigma_{s{\cal A}_s })^{-1} \vone = \frac{1}{ a + (\nu -1 ) b} \vone > 0,\]
i.e., A$2^{'}$ holds.

Note that, in the above argument, we only need $\rho = \rho_n \leq \rho_0 < 1$.  But, for the {\it irrepresentable condition} to hold, the following restriction must be in place:
\[\rho_n < \frac{1}{1+c |s_{0}|}\]
 for some constant $c$, see \cite{Zhao and Yu 2006}.  If $|s_{0}| \to \infty$, $\rho_n$ must go to zero, i.e., eventually, all the features must be statistically uncorrelated.

\vspace{0.2in}
\noindent {\bf Special case II}.  Without loss of generality, let $s_0 = \{ 1, \dots, p_0\}$.  Assume that
\begin{description}
  \item[(i)] $|\beta_1| > |\beta_2| > \cdots > |\beta_{p_0}| = Cn^{-1/2+\delta}$ for some constant C and an arbitrarily small positive $\delta$;
  \item[(ii)] The correlation matrix $\Sigma$ has the following structure:
  \[ \Sigma_{s_0 s_0} = I,  \  \  \Sigma_{j s_0} = \frac{1}{p_0} \mbox{sign}\vbeta(s_0)^\tau,  \  \mbox{for} \ j \in s_0^c.\]
\end{description}
In the following, we show that in this case the {\it irrepresentable condition} is violated but conditions A$1^{'}$-A$3^{'}$ hold, and if in addition $a2$ and $a3$ are assumed, the sequential LASSO is selection consistent.
Obviously,
\[   \Sigma_{j s_0}  \Sigma_{s_0 s_0}^{-1} \mbox{sign}\vbeta(s_0)  = 1, \]
i.e., the {\it irrepresentable  condition} does not hold.
Let $s_{*0} = \phi$. Suppose $s_{*k} = \{ 1, \dots, k\}$ for $k<p_0$.
 For any  $j\in s_{0}^c,$
\begin{eqnarray*}
\Gamma(j,s_{*k},\vbeta) & = & [(\Sigma_{js_{*k}}, \Sigma_{j s_{*k}^-}, \Sigma_{j s_{0}^c}) - \Sigma_{j s_{*k}}\Sigma_{s_{*k}s_{*k}}^{-1}(\Sigma_{s_{*k}s_{*k}}, \Sigma_{s_{*k}s_{*k}^-}, \Sigma_{s_{*k} s_{0}^c}) ]\left( \begin{array}{c}
    \vbeta(s_{*k}) \\ \vbeta(s_{*k}^-) \\  \vbeta(s_{0}^c) \end{array} \right)  \\
& =&  \Sigma_{j s_{*k}^-}  \vbeta(s_{*k}^-)  =  \sum_{j\in s_{*k}^- }|\beta_j| /p_{0} < |\beta_{k+1}| = \Gamma(k+1,s_{*k},\vbeta) \\
&=&\max_{j\in s_{*k}^- }| \Gamma(j,s_{*k},\vbeta).
\end{eqnarray*}
Thus A$1^{'}$ is satisfied. The validity of A$2^{'}$ is obvious since ${\cal A}_{s_{*k}}$ contains only one element for each $k< p_0$. A$3^{'}$ reduces to $\frac{\sqrt{n}}{\ln p_n} \min_{j \in s_0} |\beta_j| \to \infty$ which holds obviously.  $a1$ follows from (ii). Then, when $a2$ and $a3$ are also satisfied, the sequential LASSO is selection consistent.

\subsection{Sequential LASSO with EBIC as stopping rule}
\label{sec3.4}

In the previous sub sections, we have shown that, if we know $|s_0|$ a priori and stop the sequential LASSO when the number of selected features is $|s_0|$, then the set of selected features will be exactly the set of causal features with probability converging to 1.   But in practice we need a workable stopping rule for the sequential LASSO since $|s_0|$ is unknown.  The extended BIC (EBIC) proposed in \cite{Chen and Chen 2008} serves as a suitable  stopping rule because of its desirable properties of selection consistency.  The EBIC is defined as follows:
\[
\mbox{EBIC}_\gamma (s)
=
n   \ln \left( \frac{ \| \vy - H(s) \vy\|_2^2 }{ n }\right)
   +  |s| \ln n + 2 \gamma \ln { p_n \choose |s|},  \ \    \gamma \geq 0.
\]
The selection consistency of the EBIC for linear regression models  is established under different assumptions on $p_n$ and $|s_0|$ in \cite{Chen and Chen 2008} and \cite{LuoChen2011}.   The following theorem (with slight changes) is quoted from \cite{LuoChen2011}:

\begin{theorem}
Assume  model (\ref{BasicModel}) and the  condition:
\[
\lim_{n \to \infty}  \min
\{  \frac{ \Delta(s)}{ p_{0n} \ln p_n} : s_{0n} \not\subset s,  |s| \leq k_n \}  = \infty,
\]
where  $k_n = k p_{0n}$ for any fixed $k > 1$.
In  addition, assume that $p_n = O(\exp(n^\kappa))$ for  $0< \kappa <1$, $p_{0n} = O(n^{c})$,  $\min\{ |\beta_{nj}|: j \in s_{0n}\} = O(n^{-(1-b)/2})$, $0<c , \kappa <1 $, $c+\kappa < b<1$. Then,  if $\gamma > 1- \frac{\ln n}{ 2 \ln p_n}$,
\[
P \{ \min_{s: |s| \leq k_n }   \mbox{EBIC}_{\gamma} (s)
> \mbox{EBIC}_{\gamma}(s_{0n}) \} \to 1.
\]
\end{theorem}

Strictly speaking, the EBIC is to be used as a selection rule rather than a stopping rule in the procedure of sequential LASSO.  The procedure described in \S \ref{sec2} needs to be slightly modified.  In the modified procedure, instead of stopping the sequential LASSO when some stopping criterion is met,  the procedure is carried out for a specified $K$ steps, where $K$ is of order $O(n)$. Then $\mbox{EBIC}_\gamma(s_{*k}), 1 \leq k \leq K$, are computed and compared. The $s_{*k}$ that minimizes the EBIC is then selected.   It is easy to see that, if the conditions for Theorem 3.3 and Theorem 3.1 (in the case of fixed feature matrix) or Theorem 3.2 (in the case of random feature matrix) hold,  the selected set of features will be exactly the set of causal features with probability converging to 1.

In actual implementation of the procedure,  $K$ can be chosen as $rn$ for some $0<r\leq 1$.  If, for a given $r$, the minimum EBIC attains at $K$ or near $K$,  then raise $r$ to a larger value.  Eventually, the minimum EBIC will attain at a $k$ which is much less than~$K$.

Another issue is the choice of $\gamma$ in EBIC.  The selection consistency is an asymptotic property. With a finite sample, the selected set will not be exactly the same as the set of causal features.  There will be causal features that are not in the selected set. There will be also non-causal feature that are selected.  The selection accuracy is characterized by two measures,    positive discovery rate ($\mbox{PDR}_n$) and false discovery rate ($\mbox{FDR}_n$), which are defined below.  Let $s_{*}$ be the selected set of features.  Then
\[ \mbox{PDR}_n=\frac{|s_{*} \cap s_{0n}|}{|s_{0n}|},\;\; \mbox{FDR}_n=\frac{|s_{*}\cap s_{0n}^c|}{|s_{*}|}. \]
The selection consistency is equivalent to  that $\mbox{FDR}_n \to 0$ and $\mbox{PDR}_n \to 1$.
Although, the EBIC is selection consistent as long as $\gamma >  1- \frac{\ln n}{ 2 \ln p_n}$,  the convergence rate of $\mbox{PDR}_n$ and $\mbox{FDR}_n$ are different for different $\gamma$ values.    For bigger $\gamma$, the $\mbox{FDR}_n $ is smaller but the $\mbox{PDR}_n$ is also smaller.  A reasonable strategy is to maximize $\mbox{PDR}_n$  when the selection consistency is still retained.  Thus a reasonable choice is $\gamma = 1- \frac{\ln n}{ 2r \ln p_n}$ for some $r$ slightly larger than 1. In our numerical studies, we take $r= 1.5$.

\section{Numerical Study}
\label{sec4}

We report in this section our simulation study on the comparison of the sequential LASSO with FSR and ordinary LASSO.  The comparison is made in two different ways.  In the first way, the sequential LASSO and FSR are stopped at step $p_{0}$, the solution path of LASSO is computed until $p_0$ features having non-zero coefficients.  In the second way,  $p_0$ is replaced by  50 which corresponds to $0.5n$ for $n=100$,  $0.25n$ for $n=200$ and $0.1n$ for $n=500$, and the EBIC with $\gamma=1-\ln n/3\ln p_n$ is used to select the final set as described in \S \ref{sec3.4}.

The diverging pattern of $p_n$ and $p_{0n}$ are taken in consistence with the theorems on the sequential Lasso and EBIC as $( p_{0n}, p_n) = (  [4n^{0.16}], [5\exp(n^{0.3})])$. For $n=100, 200$ and 500, this yields the following table:
\begin{center}
\begin{tabular}{c|ccc} \hline
 $n$ &  100  &200 & 500   \\
   $p_n$ &   268&  672 &3,170 \\ 
$p_{0n}$ &    8  &  9   & 11    \\ \hline
\end{tabular}
\end{center}

Two types of  coefficients for causal features are considered.  For the first type, the coefficients are generated as independent random variables distributed as  $(-1)^u(4n^{-0.15}+|z|),$  where $ u\sim Bernoulli(0.4)$ and $z$ is a normal random variable
with mean $0$ and satisfies $P(|z|\geq 0.1)=0.25$.  The coefficients take both positive and negative values and are roughly of order $O(n^{-0.15})$.
For the second type,  the coefficient are generated as $2j^{0.5} n^{-0.15}$, $1\leq j\leq p_{0n}.$ The coefficients are all positive and the minimum magnitude has order $O(n^{-0.15})$ while the maximum magnitude has order $O(n^{-0.07}).$

The error variance $\sigma^2$ is determined by setting the following ratio to certain values:
\[ h =\dfrac{\vbeta^\tau \Sigma \vbeta }{\vbeta^\tau \Sigma \vbeta+\sigma^2},\]
where $\vbeta$ is the true parameter vector and $\Sigma$ is  the covariance matrix of the predictors.

Two sets of simulation study with different correlation structures of the features are considered. In the first set, the correlation structure for causal and non-causal features are not distinguished.  In the second set, different correlation structures are assumed for causal and non-causal features. The two sets of simulation study are referred to as simulation study A and simulation study B.  The correlation structures for each study are described in the following.

\vspace{0.2in}
\noindent{\bf Simulation Study A}

\begin{itemize}
\item \textbf{Structure A1:}
All the $p_n$ features are statistically independent with mean zero and variance 1.

\item\textbf{Structure A2:}
The $p_n$ features have a constant pairwise correlation, i.e., $\Sigma= (1-\rho) I+ \rho \vone \vone^\tau$, where $I$ is a $p_n$ dimensional identity matrix and $\vone$ is a $p_n$ dimensional vector of elements 1.

\item \textbf{Structure A3:} The $\Sigma$ satisfies $\Sigma_{ij}=\rho^{|i-j|}$ for all $i,j =1,2,\cdots,p_n$.
 The true features are scattered  in clusters of size 3 or 2.  
\end{itemize}

\vspace{0.2in}
\noindent{\bf Simulation Study B}

\begin{itemize}
\item \textbf{Structure B1:} Let $Z_1, \cdots, Z_{p_n}$ and ,
$W_1,\cdots,W_{p_{0n}}$  be i.i.d. random vectors with distribution $N(0, I)$.
 The feature vectors  are generated as:
\[X_j=\dfrac{Z_j+W_j}{\sqrt{2}},\;\text{for}\;j\in s_{0n};\;\;X_j=\dfrac{Z_j+\sum_{k\in s_{0n}}Z_k}{\sqrt{1+p_{0n}}}\;\text{for}\;j\notin s_{0n}.\]

\item \textbf{Structure B2:} The features in $s_{0n}$ have constant pairwise correlation.  Let $X_j, j \in s_{0n}$ be the causal feature vectors generated accordingly.  For $j \not \in s_{0n}$, the feature vectors are generated as:
\[ X_j=\vepsilon_j+\dfrac{\sum_{k\in s_{0n}}X_k}{p_{0n}}, \]
where $\vepsilon_j$'s are independent vectors from
$\N(0,0.08*\mathbf{I}_n).$ Here the variance of $\vepsilon_j$
is set to  $0.08$ in order  for the second term, which is correlated with causal features,  to dominate the variance.

\item\textbf{Structure B3:} The features are generated in the same was as in Structure B2 except that the causal features are generated according to the covariance matrix $\Sigma$ with $\Sigma_{ij}=\rho^{|i-j|} $ and $s_{0n}$ set to $\{1,2,\cdots,p_{0n}\}$.

\end{itemize}

For each setting of the simulation studies, the  PDR and FDR are averaged over 200 replicates.  For simulation study A, we only report the results with the first type of coefficients, since the results with the second type of coefficients are similar. The results of simulation study A are given in Table~\ref{table1}.  For simulation study B, the results with different types of coefficients are quite different.  The results of simulation study B are given in Table~\ref{table2} and \ref{table3} respectively for the first and second type of coefficients.  For the structures involving the parameter $\rho$, we considered two values of $\rho$: 0.3 and 0.5.   The relative performance pattern among the three methods are the same for different $\rho$ values. Only the results with $\rho=0.5$ are reported in the tables for the sake of clarity. 
The findings of the simulation studies are summarized and discussed below.

Simulation study A is discussed first. When the features are all independent (A1),  the SLasso and FSR are slightly better than Lasso in terms of both PDR and FDR.  But Lasso is still comparable with SLasso and FSR.   When the features have constant pairwise correlation (A2),  the SLasso and FSR are prominently  better than Lasso, especially when EBIC is used for feature selection, and SLasso and FSR are comparable.   When the correlations have an exponential decay (A3), again, the three methods are comparable with SLasso and FSR slightly better than Lasso.    The simulation results are quite close to each other when the known $p_{0n}$ is used as the stopping rule and when EBIC is used for selection.  Simulation study A demonstrates that, under the assumed covariance structures,  both SLasso and FSR perform better than Lasso. However,  this study does not distinguish between SLasso and FSR.

We now turn to simulation study B.  The following is the finding in the case of first type of coefficients.   When $p_{0n}$ is used as the stopping rule,  the SLasso is better than Lasso which in turn is better than FSR in terms of higher PDR and lower FDR, and the differences are quite significant. The same pattern prevails under all three correlation structures.  When EBIC is used for  selection, still, both SLasso and Lasso are much better than FSR,  the relative performances of SLasso and Lasso are reversed.  But SLasso is not much worse than Lasso.

In the case of second type of coefficients in simulation study B,  Lasso seems to be totally off target no matter whether $p_{0n}$ is used as the stopping rule or EBIC is used for selection.   SLasso is much better than FSR under all the three correlation structures when $p_{0n}$ is used as the stopping rule.  When EBIC is used for selection, SLasso is still better than FSR under correlation structure B1 and B3. Under correlation structure B2, FSR has higher PDR but also higher FDR, it is hard to say which one is better.

In general, SLasso manifests itself as the best procedure in simulation study B. This is not a coincidence.  It is due to the intrinsic property of SLasso which Lasso and FSR lack, see \S \ref{sec2}.  Under the correlation structures in simulation study B,  the non-causal features are highly correlated collectively with all the causal features and also with each other among themselves.  When causal features and/or some of non-causal features are selected, the SLasso has the capacity to resist selecting other non-causal features.

To conclude,  both simulation study A and B provide evidences for the advantage of using SLasso:  in most of cases, SLasso is the best procedure; in cases where SLasso in not the best, it is comparable with the best. In a sense, it is robust over different correlation structures of the features.

\begin{center}
\begin{table}[h]
\caption{ Results of Simulation Study A with First Type of Coefficients}
\label{table1}
\vspace{0.2in}
\begin{tabular}{c|c|ccc|ccc} \hline \hline
 \multicolumn{8}{c}{Sequential procedure stopped at step $p_{0n}$} \\ \hline
 &   & \multicolumn{3}{|c}{PDR} & \multicolumn{3}{|c}{FDR}\\ \cline{3-8}
 $n$ & Struc. & Lasso & FSR &SLasso &Lasso & FSR & SLasso \\ \hline
 100 &  A1&  .940(.084)& .990(.035)& .989(.035)&.060(.084)& .010(.035)& .011(.035)\\ 
 &A2&.642(.170) &.675(.183)& .659(.195)&.358(.171)& .325(.183) &.341(.195)\\  
 & A3 & .778(.132)& .815(.151)&.782(.169)&.222(.132)& .185(.151) &.218(.169)\\ \hline
200 & A1&   .999(.011) & 1.00(.000)& 1.00(.000)&.001(.011)& .000(.000)& .000(.000)\\ 
 &A2 &.623(.107) & .638(.129)& .638(.124)&.375(.109)& .362(.129)& .362(.124)\\ 
  & A3 &.838(.098) &.851(.112)&.840(.125)&.162(.098)&.149(.112)&.160(.125)\\
 \hline
500&A1&1.00(.000)& 1.00(.000)& 1.00(.000)&.000(.000)& .000(.000) &.000(.000)\\ 
  & A2 & .732(.142)& .736(.144)& .736(.144)& .268(.142)& .264(.144)& .264(.144)\\ 
 &A3&.893(.089)& .893(.098)& .891(.103)& .107(.089)& .107(.098)& .109(.103)\\     \hline \hline
 \multicolumn{8}{c}{Final set selected by EBIC} \\ \hline
 &  & \multicolumn{3}{|c}{PDR} & \multicolumn{3}{|c}{FDR}\\ \cline{3-8}
 $n$ & Struc. & Lasso & FSR &SLasso &Lasso & FSR & SLasso \\ \hline
100 & A1&  .989(.053)& .992(.031)& .992(.031)&.071(.099)& .049(.080) &.051(.083)\\ 
  &A2 &.472(.282)& .503(.283)& .496(.281)&.108(.189)& .076(.169)& .088(.184)\\ 
  &A3&.768(.198) &.799(.174) &.783(.188)&.104(.133) &.101(.128)& .118(.138)\\ \hline
200 &A1& 1.00(.000) &1.00(.000)& 1.00(.000)&.011(.033)& .029(.053)& .029(.053)\\ 
  & A2 &.403(.221)  & .417(.232)& .417(.229)&.044(.108)& .011(.047)& .014(.050)\\ 
  &A3  &.839(.122)& .839(.128) & .836(.132)&.070(.113) &.058(.087) & .064(.093)\\ \hline
500&A1& 1.00(0.00)& 1.00(.000) &1.00(.000)& .013(.036)& .017(.038)& .017(.038) \\ 
  & A2 & .679(.205)& .684(.202)& .682(.203)& .083(.128)& .112(.137)& .111(.137) \\ 
 &A3 &.894(.097)& .888(.106)& .888(.106)& .029(.060)& .026(.056)& .027(.056)\\  \hline \hline
 \end{tabular}
\end{table}
\end{center}
\vspace{1in}

\newpage
\begin{center}
\begin{table}[t]
\caption{ Results of Simulation Study B with First Type of Coefficients }
\label{table2}
\vspace{0.2in}
\begin{tabular}{c|c|ccc|ccc} \hline \hline
  \multicolumn{8}{c}{Sequential procedure stopped at step $p_{0n}$} \\ \hline
 &   & \multicolumn{3}{|c}{PDR} & \multicolumn{3}{|c}{FDR}\\ \cline{3-8}
 $n$ & Struc.& Lasso & FSR &SLasso &Lasso & FSR & SLasso \\ \hline
100&B1& .682(.368)& .477(.276)& .728(.260)& .430(.325)& .523(.276)& .456(.213) \\ 
  & B2& .778(.301)& .628(.366)& .924(.126) &.268(.278)& .372(.366)& .228(.155)\\  
 &B3&.783(.302)& .645(.336)& .902(.164)&.254(.282)& .355(.336)& .227(.181)\\  \hline
200&B1& .665(.343)& .508(.285)& .729(.244)& .497(.280)& .492(.285)& .506(.187) \\ 
  & B2&.727(.314)& .569(.366)& .939(.104)&.305(.303)& .431(.366)& .248(.148) \\  
 &B3 &.788(.303)& .689(.360)& .909(.155)&.245(.290)& .311(.360)& .268(.185) \\ \hline
 500&B1& .686(.358)& .478(.263)& .756(.249)& .552(.264)& .522(.263)& .590(.158) \\ 
  & B2&  .694(.381) & .644(.376) & .963(.103)& .421(.322) & .356(.376) & .321(.156)\\  
 &B3& .766(.339) & .726(.362) & .917(.163) &.285(.316) & .273(.362) & .230(.205)\\ \hline \hline
\multicolumn{8}{c}{Final set selected by EBIC} \\ \hline
 &  & \multicolumn{3}{|c}{PDR} & \multicolumn{3}{|c}{FDR}\\ \cline{3-8}
 $n$ & Struc. & Lasso & FSR &SLasso &Lasso & FSR & SLasso \\ \hline
100&B1& .685(.372)& .532(.262)& .751(.319)& .328(.322)& .816(.179)& .727(.232)\\ 
  &B2& .853(.349)& .656(.345)& .647(.393) &.188(.340)& .789(.195)& .291(.246) \\ 
 &B3 & .859(.259)& .661(.327)& .728(.378) &.198(.266)& .766(.218)& .269(.253) \\  \hline
200&B1&.656(.344)& .505(.288)& .623(.334)&.356(.319)& .488(.230)& .390(.275)\\ 
  &B2& .908(.287)& .563(.372)& .604(.377) &.121(.285)& .367(.254)& .262(.208) \\  
 &B3&.874(.244)& .686(.365)& .768(.349) &.178(.270)& .307(.258)& .216(.228)\\  \hline
500&B1& .680(.362)& .476(.265)& .643(.350)& .330(.356)& .535(.217)& .365(.322)\\ 
     & B2&  .949(.214) & .642(.378) & .766(.359) & .075(.219) & .312(.285) & .196(.223) \\   
 &B3&.912(.215) & .724(.365) & .817(.310)&.151(.279) & .270(.290) & .183(.230)\\ \hline \hline
 \end{tabular}
\end{table}
\end{center}
\vspace{1in}

\newpage
\begin{center}
\begin{table}[h]
\caption{ Results of Simulation Study B with Second Type of Coefficients }
\label{table3}
\vspace{0.2in}
\begin{tabular}{c|c|ccc|ccc} \hline \hline
  \multicolumn{8}{c}{Sequential procedure stopped at step $p_{0n}$} \\ \hline
 &   & \multicolumn{3}{|c}{PDR} & \multicolumn{3}{|c}{FDR}\\ \cline{3-8}
 $n$ & Struc. & Lasso & FSR &SLasso &Lasso & FSR & SLasso \\ \hline
100&B1&.004(.029)& .366(.115)& .716(.123)&.997(.021)& .634(.115)& .440(.145) \\
  &B2& .153(.141)& .133(.100)& .744(.127)&.856(.137)& .867(.100)& .380(.129)\\ 
 &B3& .163(.136)& .295(.087)& .725(.123)&.852(.125)& .705(.087)& .391(.131) \\  \hline
200&B1& .000(.000)& .432(.093)& .776(.103)& 1.00(.000)& .568(.093)& .423(.135) \\ 
  & B2& .064(.110)& .130(.098)& .793(.125) &.948(.090)& .870(.098)& .368(.133) \\  
 &B3&.079(.107)& .360(.077)& .740(.120)&.937(.085)& .640(.077)& .425(.151) \\  \hline
500&B1&.000(.000) & .485(.072) & .870(.102)& 1.00(.000) & .516(.072) & .376(.157) \\ 
  & B2&.002(.014) & .176(.093) & .826(.116)&.999(.010) & .825(.093) & .399(.146)\\  
 &B3& .010(.038) & .447(.068) & .836(.111) &.994(.023) & .553(.068) & .398(.172) \\  \hline
 \hline
\multicolumn{8}{c}{Final set selected by EBIC} \\ \hline
 & & \multicolumn{3}{|c}{PDR} & \multicolumn{3}{|c}{FDR}\\ \cline{3-8}
 $n$ & Struc.& Lasso & FSR &SLasso &Lasso & FSR & SLasso \\ \hline
100&B1& .002(.035)& .449(.168)& .619(.319)&.998(.024)& .826(.151)& .712(.226)\\ 
  & B2& .000(.000)& .116(.161)& .018(.045)&1.00(.000)& .955(.110)& .944(.148)\\  
 &B3& .004(.033)& .305(.159)& .197(.150)&.994(.047)& .757(.225)& .528(.170) \\  \hline
200&B1& .000(.000)& .390(.113)& .533(.199)&1.00(.000)& .565(.090)& .422(.126)\\
  & B2&  .000(.000)& .021(.054)& .026(.058)&1.00(.000)& .925(.178)& .925(.163)\\  
 &B3& .004(.039)& .277(.074)& .273(.092) &.997(.033)& .460(.100)& .428(.130)\\ \hline
500&B1& .000(.000) & .530(.083) & .758(.198)& 1.00(.000) & .529(.062) & .363(.136)\\ 
  & B2& .000(.000) & .050(.063) & .049(.065)&1.00(.000) & .844(.187) & .863(.182) \\  
 &B3&.003(.042) & .381(.066) & .398(.097)&.999(.018) & .434(.062) & .379(.126) \\ 
 \hline \hline
 \end{tabular}
\vspace{3.2in}
\end{table}

\end{center}

\newpage
\end{document}